# Thermal processes induced in carbon nanotubes by attosecond laser pulses

Janina Marciak-Kozłowska, Mirosław Kozłowski*

Institute of Electron Technology, Al. Lotników 32/46, 02-668 Warszawa, Poland

* Corresponding author

## Abstract

In this paper the heat transport in carbon nanotubes is investigated. When the dimension of the structure is of the order of the de Broglie wave length the transport phenomena must be analyzed within quantum mechanics. In this paper we developed the Dirac type thermal equation .The solution of the equation-the temperature fields for electrons can be damped or can oscillate depending on the dynamics of the scattering.

## 1. Introduction

The exceptional properties of carbon nanotubes (CNTs), including ballistic transport and semiconducting behavior with band-gaps in the range of 1 eV, have sparked a large number of theoretical [1-3]and experimenta [4-6] studies. The possibility to use CNTs to replace crystalline silicon as the basis for high-performance transistors has prompted significant effort to reduce the size of CNT field-effect transistors (CNTFETs) in an effort to understand their scaling behavior and ultimate limits. In this context we present CNT transistors with channel lengths below 20 nm, which have characteristics comparable to those of much larger silicon-based demonstrators with similar channel lengths.[7]

Since the first CNTFET was demonstrated in 1998,[9] their characteristics have been continuously and rapidly improved, particularly in the past few years.[5,6] One critical aspect is the optimization of the source and drain contacts to minimize the Schottky barrier (SB) due to the mismatch between the CNT and contact metal work functions. As discussed by Guo et al.,[1] the formation of significant SBs of the order of 0.2 eV can severely affect the function of short-channel CNTFETs because electron injection through the SBs at higher drain-source biases reduces the on/off current ratio. In addition, substantial SBs reduce the on-current, as shown by the difference between Ti- and Pd-contacted CNTFETs.[5,6] A further important aspect is the band gap of the CNTs, which scale as $0.9/d$ eV, where $d$ is the nanotube diameter in nanometers.[10] As shown by Javey et al.,[11] it is possible to form almost perfect ohmically contacted CNTFETs with 1.5-2 nm diameter nanotubes using Pd contacts. These devices have on-conductivities close to the quantum conduction limit $2G_0 = 155$ µS. However, these CNTs have band gaps of only 0.4-0.5 eV, leading to high off-currents for short-channel devices [5] Therefore, we have concentrated on CNTs with smaller diameters of around 0.7-1.1 nm that have larger band gaps of around 0.8-1.3 eV and are suitable for short-channel devices. [12] Catalytic chemical vapor deposition (CCVD) provides the possibility to selectively and cleanly grow CNTs with a narrow diameter distribution for the production of significant numbers of device demonstrators[12] or high current transistors.[13] In a recent publication we have demonstrated the possibility to grow small diameter CNTs for long channel CNTFETs gated using the substrate or e-beam defined top gates.[14]

In this work we have investigated the thermal processes in carbon nanotubes with channel lengths below 20 nm. [15]

The breakthrough progress has been made recently in generation and detection of ultrashort, attosecond laser pulses with high harmonic generation technique [16,17]. This is the beginning of the attophysics age in which many-electron dynamics will be investigated in real time.

In the existing new laser projects [18] the generation of 100 GW-level attosecond X-ray pulses is investigated. With ultra-short (attosecond) high energetic laser pulses the relativistic multi-electrons states can be generated.

For relativistic one electron state the Dirac equation is the master equation. In this paper we develop and solve the Dirac type thermal equation for multi-electron states generated in laser interaction with matter. In this paper the Dirac one dimensional thermal equation will be applied to study of the generation of the positron-electron pairs. It will be shown that the cross section is equal to the Thomson cross-section for the electron-electron scattering.

## 2. Derivation of the 1+1 dimensional Dirac equation for thermal processes

As pointed in paper [19] spin-flip occurs only when there is more than one dimension in space. Repeating the discussion of deriving the Dirac equation [19] for the case of one spatial dimension, one easily finds that the Dirac matrices *α* and *β* are reduced to 2x2 matrices that can be represented by the Pauli matrices [19]. This fact simply implies that if there is only one spatial dimension, there is no spin. It should be instructive to show explicitly how to derive the 1+1 dimensional Dirac equation.

As discussed in textbooks [19,20] a wave equation that satisfies relativistic covariance in space-time as well as the probabilistic interpretation should have the form:

$$i\hbar \frac{\partial}{\partial t}\Psi(x,t) = \left[c\boldsymbol{\alpha}\left(-i\hbar \frac{\partial}{\partial x}\right) + \boldsymbol{\beta} m_0 c^2\right]\Psi(x,t). \qquad (1)$$

To obtain the relativistic energy-momentum relation $E^2 = (pc)^2 + m_0^2 c^4$ we postulate that (1) coincides with the Klein-Gordon equation

$$\left[\frac{\partial^2}{\partial (ct)^2} - \frac{\partial^2}{\partial x^2} + \left(\frac{m_0 c}{\hbar}\right)^2\right]\Psi(x,t) = 0. \qquad (2)$$

By comparing (1) and (2) it is easily seen that $\boldsymbol{\alpha}$ and $\boldsymbol{\beta}$ must satisfy

$$\boldsymbol{\alpha}^2 - \boldsymbol{\beta}^2 = 1, \qquad \boldsymbol{\alpha}\boldsymbol{\beta} + \boldsymbol{\beta}\boldsymbol{\alpha} = 0. \qquad (3)$$

Any two of the Pauli matrices can satisfy these relations. Therefore, we may choose $\boldsymbol{\alpha} = \sigma_x$ and $\boldsymbol{\beta} = \sigma_z$ and we obtain:

$$i\hbar \frac{\partial}{\partial t}\Psi(x,t) = \left[c\sigma_x\left(-i\hbar \frac{\partial}{\partial x}\right) + \sigma_z m_0 c^2\right]\Psi(x,t), \qquad (4)$$

where $\Psi(x,t)$ is a 2-component spinor.

The Eq. (4) is the Weyl representation of the Dirac equation. We perform a phase transformation on $\Psi(x,t)$ letting $u(x,t) = \exp\left(\frac{imc^2 t}{\hbar}\right)\Psi(x,t)$. Call $u$'s upper (respectively, lower) component $u_+(x,t)$, $u_-(x,t)$; it follows from (4) that $u_\pm$ satisfies

$$\frac{\partial u_\pm(x,t)}{\partial t} = \pm c \frac{\partial u_\pm}{\partial x} + \frac{im_0 c^2}{\hbar}(u_\pm - u_\mp). \qquad (5)$$

Following the physical interpretation of the equation (5) it describes the relativistic particle (mass $m_0$) propagates at the speed of light $c$ and with a certain *chirality* (like a two component neutrino) except that at random times it flips both direction of propagation (by $180^0$) and chirality.

In monograph [21] we considered a particle moving on the line with fixed speed $w$ and supposed that from time to time it suffers a complete reversal of direction, $u(x,t) \Leftrightarrow v(x,t)$, where $u(x,t)$ denotes the expected density of particles at $x$ and at time $t$ moving to the right, and $v(x,t) \equiv$ expected density of particles at $x$ and at time $t$ moving to the left. In the following we perform the change of the abbreviation

$$\begin{aligned} u(x,t) &\to u_+, \\ v(x,t) &\to u_-. \end{aligned} \qquad (6)$$

Following the results of the paper [6] we obtain for the $u_\pm(x,t)$ the following equations

$$\begin{aligned} \frac{\partial u_+}{\partial t} &= -w\frac{\partial u_+}{\partial x} - \frac{w}{\lambda}\big((1-k)u_+ - ku_-\big), \\ \frac{\partial u_-}{\partial t} &= w\frac{\partial u_-}{\partial x} + \frac{w}{\lambda}\big(ku_+ + (k-1)u_-\big). \end{aligned} \qquad (7)$$

In equation (7) $k(x)$ denotes the number of the particles which are moving in left (right) direction after the scattering at $x$. The mean free path for scattering is equal $\lambda$, $\lambda = w\tau$, where $\tau$ is the relaxation time for scattering.

Comparing formulae (5) and (7) we conclude that the shapes of both equations are the same. In the subsequent we will call the set of the equations (7) *the Dirac equation* for the particles with velocity $w$, mean free path $\lambda$.

For thermal processes we define $T_{+,-} \equiv$ the temperature of the particles with chiralities + and − respectively and with analogy to equation (7) we obtain:

$$\begin{aligned} \frac{\partial T_+}{\partial t} &= -w\frac{\partial T_+}{\partial x} - \frac{w}{\lambda}\big((1-k)T_+ - kT_-\big), \\ \frac{\partial T_-}{\partial t} &= w\frac{\partial T_-}{\partial x} + \frac{w}{\lambda}\big(kT_+ + (k-1)T_-\big), \end{aligned} \qquad (8)$$

where $\frac{w}{\lambda} = \frac{1}{\tau}$.

In one dimensional case we introduce one dimensional cross section for scattering

$$\sigma(x,t) = \frac{1}{\lambda(x,t)}. \tag{9}$$

## 3. The solution of the Dirac equation for stationary temperatures in one carbon nanotubes

In the stationary state thermal transport phenomena $\frac{\partial T_{+,-}}{\partial t} = 0$ and Eq. (8) can be written as

$$\frac{dT_+}{dx} = -\sigma\big((1-k)T_+ + kT_-\big),$$
$$\frac{dT_-}{dx} = \sigma(k-1)T_- + \sigma k T_+. \tag{10}$$

After the differentiation of the equation (9) we obtain for $T_+(x)$

$$\frac{d^2 T_+}{dx^2} - \frac{1}{\sigma k}\frac{d}{dx}(\sigma k)\frac{dT_+}{dx} + T_+\left[\sigma^2(2k-1) + \frac{d\sigma}{dx}(1-k) + \frac{\sigma(k-1)}{\sigma k}\frac{d(\sigma k)}{dx}\right] = 0.$$

Equation (10) can be written in a compact form

$$\frac{d^2 T_+}{dx^2} + f(x)\frac{dT_+}{dx} + g(x)T_+ = 0,$$

where

$$f(x) = -\frac{1}{\sigma}\left(\frac{\sigma}{k}\frac{dk}{dx} + \frac{d\sigma}{dx}\right),$$
$$g(x) = \sigma^2(x)(2k-1) - \frac{\sigma}{k}\frac{dk}{dx}. \tag{11}$$

In the case for constant $\frac{dk}{dx} = 0$ we obtain

$$f(x) = -\frac{1}{\sigma}\frac{d\sigma}{dx},$$
$$g(x) = \sigma^2(x)(2k-1). \tag{12}$$

With functions $f(x)$, $g(x)$ described by formula (12) the general solution of Eq. (12) has the form:

$$T_+(x) = C_1 e^{(1-2k)^{\frac{1}{2}}\int \sigma(x)dx} + C_2 e^{-(1-2k)^{\frac{1}{2}}\int \sigma(x)dx} \tag{13}$$

and

$$T_-(x) = \frac{\left[(1-k)+(1-2k)^{\frac{1}{2}}\right]}{k} \times$$
$$\left[C_1 e^{(1-2k)^{\frac{1}{2}}\int\sigma(x)dx} + \frac{(1-k)-(1-2k)^{\frac{1}{2}}}{(1-k)+(1-2k)^{\frac{1}{2}}} C_2 e^{-(1-2k)^{\frac{1}{2}}\int\sigma(x)dx}\right]. \tag{14}$$

The formulae (13) and (14) describe three different mode for heat transport. For $k = \frac{1}{2}$ we obtain $T_+(x) = T_-(x)$ while for $k > \frac{1}{2}$, i.e. for heat carrier generation $T_+(x)$ and $T_-(x)$ oscillate for $(1-2k)^{\frac{1}{2}}$ is a complex number. For $k < \frac{1}{2}$ i.e. for absorption $T_+(x)$ and $T_-(x)$ decrease as the function of $x$.

In the subsequent we will consider the solution of Eq. (9) for Cauchy conditions:
$$T_+(0) = T_0, \quad T_-(a) = 0. \tag{15}$$

Boundary conditions (15) describes the generation of heat carriers by illuminating the left end of one dimensional slab (with length $a$) by laser pulse. From formulae (13) and (14) we obtain:

$$T_+(x) = \frac{2T_0 e^{[f(0)-f(a)]}}{1+\beta e^{2[f(0)-f(a)]}} \times \frac{(1-2k)^{\frac{1}{2}}\cosh[f(x)-f(a)]+(k-1)\sinh[f(x)-f(a)]}{(1-2k)^{\frac{1}{2}}-(k-1)}, \tag{16}$$

$$T_-(x) = \frac{2T_0 e^{2[f(0)-f(a)]}\left[(k-1)+(1-2k)^{\frac{1}{2}}\right]\sinh[f(x)-f(a)]}{(1+\beta e^{-2[f(a)-f(0)]})k}. \tag{17}$$

In formulae (16) and (17)
$$\beta = \frac{(1-2k)^{\frac{1}{2}}+(k-1)}{(1-2k)^{\frac{1}{2}}-(k-1)} \tag{18}$$

and
$$f(x) = (1-2k)^{\frac{1}{2}}\int\sigma(x)dx,$$
$$f(0) = (1-2k)^{\frac{1}{2}}\left[\int\sigma(x)dx\right]_0, \tag{19}$$
$$f(a) = (1-2k)^{\frac{1}{2}}\left[\int\sigma(x)dx\right]_a.$$

With formulae (16) and (17) for $T_+(x)$ and $T_-(x)$ we define the asymmetry $A(x)$ of the temperature $T(x)$

$$A(x) = \frac{T_+(x) - T_-(x)}{T_+(x) + T_-(x)}, \tag{20}$$

$$A(x) = \frac{\dfrac{(1-2k)^{\frac{1}{2}}}{(1-2k)^{\frac{1}{2}} - (k-1)}\cosh[f(x) - f(a)] - \dfrac{1-2k}{(1-2k)^{\frac{1}{2}} - (k-1)}\sinh[f(x) - f(a)]}{\dfrac{(1-2k)^{\frac{1}{2}}}{(1-2k)^{\frac{1}{2}} - (k-1)}\cosh[f(x) - f(a)] - \dfrac{1}{(1-2k)^{\frac{1}{2}} - (k-1)}\sinh[f(x) - f(a)]} \tag{21}$$

From formula (21) we conclude that for elastic scattering, i.e. when $k = \dfrac{1}{2}$, $A(x) = 0$, and for $k \neq \dfrac{1}{2}$, $A(x) \neq 0$.

In the monograph [20] we introduced the relaxation time $\tau$ for quantum heat transport

$$\tau = \frac{\hbar}{mv^2}. \tag{22}$$

In formula (22) $m$ denotes the mass of heat carriers electrons and $v = \alpha c$, where $\alpha$ is the fine structure constant for electromagnetic interactions. As was shown in monograph [21], $\tau$ is also the lifetime for positron-electron pairs in vacuum. When the duration time of the laser pulse is shorter than $\tau$, then to describe the transport phenomena we must use the hyperbolic transport equation. Recently the structure of water was investigated with the attosecond $(10^{-18}\,\text{s})$ resolution [22]. Considering that $\tau \approx 10^{-17}$ s we argue that to study performed in [22] open the new field for investigation of laser pulse with matter. In order to apply the equations (9) to attosecond laser induced phenomena we must know the cross section $\sigma(x)$. Considering formulae (9) and (22) we obtain

$$\sigma(x) = \frac{mv}{\hbar} = \frac{me^2}{\hbar^2} \tag{23}$$

and it occurs $\sigma(x)$ is the Thomson cross section for electron-electron scattering. With formula (23) the solution of Cauchy problem has the form:

$$T_+(x) = \frac{2T_0 e^{-(1-2k)^{\frac{1}{2}}\frac{me^2}{\hbar^2}a}}{\left[1+\beta e^{-2(1-2k)^{\frac{1}{2}}\frac{me^2}{\hbar^2}a}\right]} \times$$

$$\frac{(1-2k)^{\frac{1}{2}}\cosh\left[(1-2k)^{\frac{1}{2}}\frac{me^2}{\hbar^2}(x-a)\right]+(k-1)\sinh\left[(1-2k)^{\frac{1}{2}}\frac{me^2}{\hbar^2}(x-a)\right]}{(1-2k)^{\frac{1}{2}}-(k-1)}, \quad (24)$$

$$T_-(x) = \frac{2T_0 e^{-\frac{(1-2k)^{\frac{1}{2}}me^2 a}{\hbar^2}}\left[(k-1)-(1-2k)^{\frac{1}{2}}\right]}{\left(1+\beta e^{-2(1-2k)^{\frac{1}{2}}\frac{me^2}{\hbar^2}a}\right)k} \times$$

$$\sinh\left[(1-2k)^{\frac{1}{2}}\frac{me^2}{\hbar^2}(x-a)\right].$$

## 4. Conclusions

In this paper the one dimensional Dirac type thermal equation for carbon nanotubes was developed and solved. It was shown that depending on the dynamics of the heat carriers scattering the damped or oscillated temperature field can be generated in carbon nanotubes. When the laser pulse generates relativistic electrons the cross section for the generation of electron-positron pairs is equal to the Thomson cross section.